\documentclass[12pt]{article}

\usepackage{epsf}
\usepackage[dvips]{graphicx}

\newcommand\ba{\begin{eqnarray}}
\newcommand\ea{\end{eqnarray}}
\newcommand\be{\begin{equation}}
\newcommand\ee{\end{equation}}
\newcommand{\br}[1]{\left( #1 \right)}
\newcommand{\brs}[1]{\left[ #1 \right]}

\newcommand{\brm}[1]{\left| #1 \right|}
\begin{document}

\begin{center}
{\bfseries EXCLUSIVE PROCESSES INDUCED BY ANTIPROTONS: OPPORTUNITIES FOR QCD STUDIES }
\vskip 5mm
E. Tomasi-Gustafsson$^{(1)}$, E.~A.~Kuraev$^{(2)}$, Yu.~M.~Bystritskiy$^{(2)}$
\vskip 5mm

{\small
{\it
(1) CEA,IRFU,SPhN, Saclay,  and
Univ Paris-Sud, CNRS/IN2P3, IPN Orsay, UMR 8608, France}
}
\\

{\small
{\it
(2) Joint Institute for Nuclear Research, BLTP, 91406 Dubna, Russia}
}
\end{center}

\vskip 5mm

\begin{center}
\begin{minipage}{150mm}
\centerline{\bf Abstract}
The detection of exclusive processes induced by antiproton beams in the GeV range at PANDA(FAIR) opens possibilities for testing QCD predictions on the nucleon structure. In particular, the measurement of electromagnetic nucleon form factors through the proton-antiproton annihilation in a lepton pair can be studied. However, other processes, although interesting by themselves, play the role of background. We will focus on two examples: an 'electromagnetic process', the S-channel annihilation of proton and antiproton into a neutral pion and a virtual photon (followed by lepton pair emission) and a possible mechanism for the production of kaon and pion pairs, through physical vacuum excitations.

\end{minipage}
\end{center}
\section{Introduction}
Antiproton reactions are considered a privileged tool for QCD studies, due to the large gluonic content involved in the interaction with other hadrons. Annihilation processes give unique information on the $q\bar q$ interaction. Electromagnetic final channels, such as $\bar p+p\to e^++e^-$ are described in terms of time-like form factors, which are complex functions of one variable, the total energy squared $s=q^2$. The accessible kinematical region corresponds to $s\ge 4M^2=3.5$ GeV$^2$ ($M$ is the proton mass). It has been suggested in Ref. \cite{Re65} that one may reach smaller $s$ values, (the 'unphysical region'), with three-body reactions, such as  $\bar p+p\to e^++e^-+\pi^0$, which sensitivity to nucleon electromagnetic and axial FFs has been previously studied \cite{Ad07}, through a mechanism denoted the 'scattering channel exchange' (Fig. 1a,b). In this talk we discuss another reaction mechanism, the 'annihilation channel exchange' through s-channel vector meson exchange (Fig.1c). This one can be considered a physical background, which relative contribution may be evaluated as follows. The cross section for 'annihilation' channel is $d\sigma_{a}\propto 1/s$ while, extrapolating the Regge behavior,  $d\sigma_{s}\propto \frac{1}{q^2}\left( \frac{s}{M^2}\right )^{2[\alpha_p(q^2)-1]}$. Here $q^2$ is the transferred momentum, $\alpha_p(q^2)$ is the Regge trajectory of the proton: $\alpha_p(q^2)<1/2$ for negative values of $q^2$. At large scattering angles $q^2\sim s$, therefore one expects a small ratio:
\ba
\frac{d\sigma_{s}}{d\sigma_{a}}\propto
\br{\frac{s}{M^2}}^{2[\alpha_p(q^2)-1]} \ll 1.
\label{eq:eq2}
\ea
We calculate the s-channel differential cross section, in the framework of the Born approximation, to be compared with previous estimates of the 'scattering' channel.

The $\bar p p$ annihilation into a hadrons has a much larger cross section than the annihilation into electromagnetic channels. We suggest a particular mechanism of production of  kaon and pion pairs, in the threshold region, where $L=0$, through excitation of physical vacuum. The meson production in $\bar p p$ annihilation occurs through the rearrangement of the constituent quarks. Selection rules require that $J=1$, so the $\pi+\pi^-$ final channel originates from the $^3S_1$ state. As the strange quark content of the proton is very small, kaon production is forbidden through such a mechanism. However, a kaon pair can be produced from the excited vacuum. In this case, a singlet state, $^1S_0$ can produce any pair of current quarks ($u\bar u$, $d\bar d$ or $s\bar s$ with equal probability), which, after interacting, become observable as mesons \cite{EK10}. 

\section{The reaction $\bar p+p\to \gamma^* + \pi^0$} 
\begin{figure}
\begin{center}
\mbox{\epsfxsize=16.cm\leavevmode \epsffile{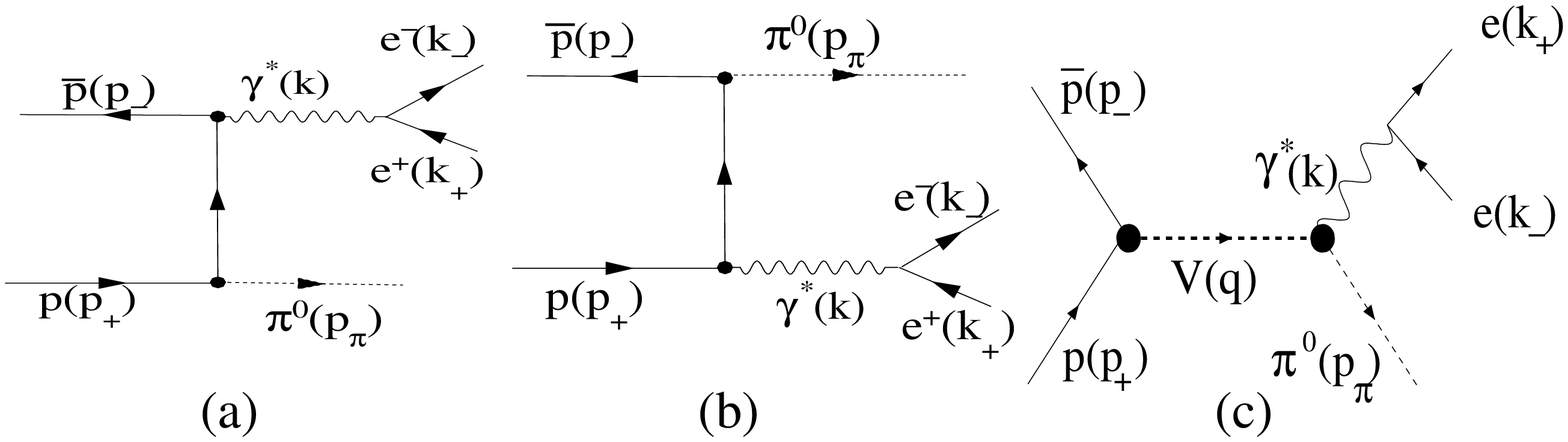}}
\caption{Feynman diagram for $\bar p+p \to e^++e^-+ \pi^0$
(a) and (b) in $t$-channel ('scattering'); (c) in $s$-channel ('annihilation').}
\label{Fig:Fig1}
\end{center}
\end{figure}

Let us consider the reaction (Fig.\ref{Fig:Fig1}):
\ba
\bar p\br{p_-} + p\br{p_+}
\to
\gamma^*\br{k} + \pi^0\br{p_\pi}
\to
e^+\br{k_+} + e^-\br{k_-} + \pi^0\br{p_\pi},
\label{eq:eqPPM}
\ea
where the four momenta of the particles are indicated in parentheses. Figs.
 \ref{Fig:Fig1}a,b correspond to the 'scattering channel' which contains information on the electromagnetic proton FFs, whereas  Fig. \ref{Fig:Fig1}c  illustrates the s-channel exchange of a vector meson $V$, $V=\rho$, $\omega$, $\phi$, of interest here.

The general expression of the matrix element, corresponding to
the diagram of Fig. \ref{Fig:Fig1}c, is:
\ba
{\cal M}=4\pi\alpha\frac{G_{V\pi\gamma^*}}{e}
\frac{\epsilon_{\mu\nu\rho\sigma }q^{\rho}k^{\sigma}}
{k^2\br{q^2-M^2_{V}+iM_{V}\Gamma_{V}}}{\cal J}_p^{\mu}{\cal J}_e^{\nu},
\label{eq:Matrix}
\ea
where $e=\sqrt{4\pi \alpha}$ is the elementary electric charge
($\alpha \approx 1/137$), $M_{V}$ and $\Gamma_{V}$ are the mass and the
total width of intermediate vector meson $V$. The
lepton electromagnetic current and the current related to the $Vp\bar p$ vertex have a standard form
\ba
{\cal J}_e^{\nu}=\bar u(k_-)\gamma^{\nu}v(k_+),~{\cal J}_p^{\mu}=\bar v(p_-) \Gamma_V^{\mu} u(p_+), 
\label{eq:LeptonCurrent}
\ea
with
\be
\Gamma_V^{\mu}=F_1^V\br{q^2}\gamma^{\mu}+
\displaystyle\frac {\sigma^{\mu\nu}q_{\nu}}{2M}F_2^V\br{q^2}
=
\left [F_1^V\br{q^2}+F_2^V\br{q^2}\right ]\gamma^{\mu}+ \frac {\Delta^{\mu}}{2M}F_2^V\br{q^2},
\label{eq:VppVertexes}
\ee
where $\sigma^{\mu\nu}=\br{-1/2}\br{\gamma^\mu\gamma^\nu-\gamma^\nu\gamma^\mu}$, and $\Delta=p_--p_+$.
In principle, all vector mesons may contribute to the intermediate
state. However, the contribution of intermediate $\omega$-meson is expected to be dominant.
\newpage

\begin{figure}[htb]
\begin{minipage}[t]{80mm}
\includegraphics[width=8cm]{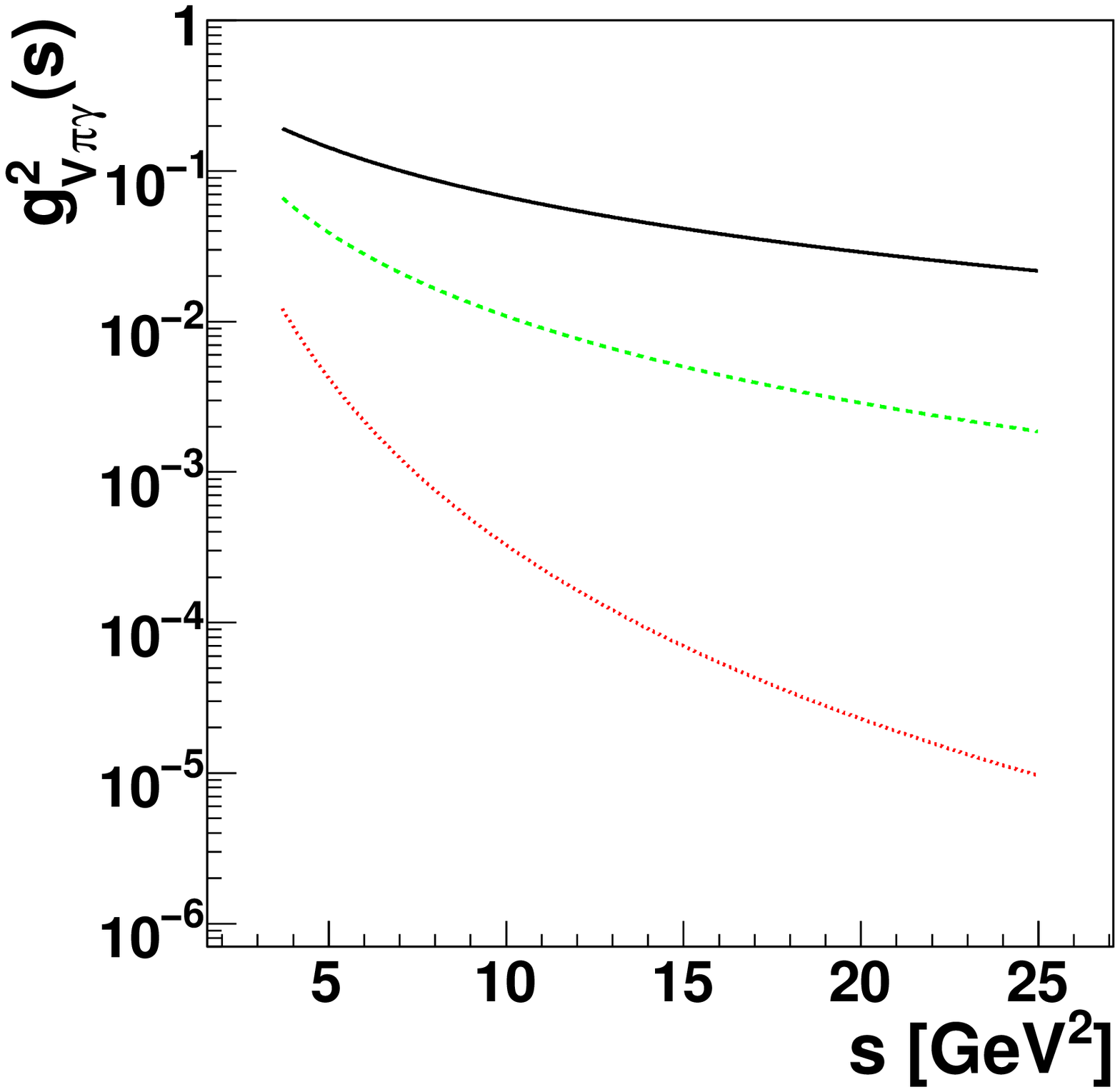}
\caption{ $G_{V\pi\gamma^*}$ as function of $q^2$ from  Nambu-Jona-Lasinio model (black, solid line), monopole (dashed line) and dipole (dotted line) dependences.}
\label{Fig:coupling}
\end{minipage}
\hspace{\fill}
\begin{minipage}[t]{80mm}
\includegraphics[width=8cm]{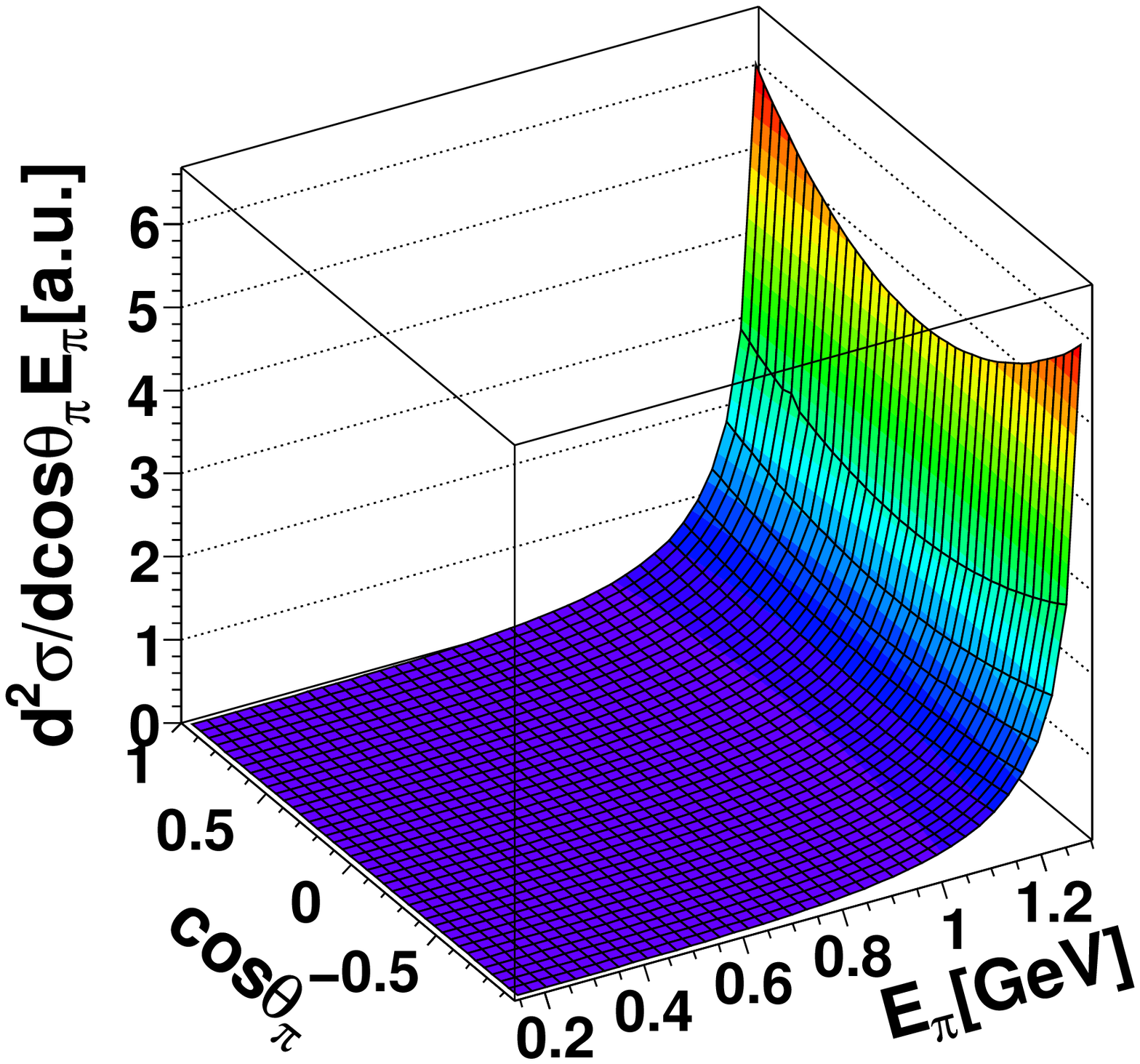}
\caption{Bidimensional plot of the cross section as a function of the pion CMS energy and angle, at fixed total energy $s=$ 7 GeV.}
\label{Fig:bidim}
\end{minipage}
\end{figure}

For $\omega$-exchange, one finds the following expression for the double differential cross section:
\ba
\frac{d^2\sigma}{dE_{\pi}d\cos\theta_{\pi}}=
\sigma_0\frac{E^3_{\pi}\br{2-\beta^2\sin^2\theta_{\pi}}}
{s\br{s+M_\pi^2-2E_{\pi}\sqrt{s}}},
\label{eq:eq2a}
\ea
where $\cos\theta_{\pi}=\cos\theta_{\pi}$, $\beta=\sqrt{1-(4M^2)/s}$ and  the coefficient $\sigma_0$ which has the dimension
of cross section is factorized out:
\ba
    \sigma_0
    =
    \frac{G^2_{\omega\pi\gamma^*}\br{s}
        \alpha
        \brm{F_1^\omega\br{s}}^2 s^2
    }
    {
     2^6\pi^3 3 \beta~
        \brs{\br{s-M^2_\omega}^2+M_\omega^2\Gamma_\omega^2}
    }.
\ea
The energy distribution of the final pion is:
\ba
\frac{d\sigma}{dE_{\pi}}
&=&
\int\limits_{-1}^1 dc_\pi
\frac{d^2\sigma}{dE_{\pi}d\cos\theta_{\pi}}
=
\sigma_0
\frac{4E^3_{\pi}}
{s(s+M^2_{\pi}-2E_{\pi}\sqrt{s})}
\br{1-\displaystyle\frac{\beta^2}{3}}.
\label{eq:EnergySpectrum}
\ea
The angular distribution has the form
\ba
\frac{d\sigma}{d\cos\theta_{\pi}}
=
\int\limits_{M_\pi}^{\frac{\sqrt{s}}{2}\br{1 + \frac{M_\pi^2}{s} - \frac{4 m_e^2}{s}}} dE_\pi
\frac{d^2\sigma}{dE_{\pi}d\cos\theta_{\pi}}
=
\frac{\sigma_0}{16}
\br{2-\beta^2\sin^2\theta_{\pi}}
\br{\ln \frac{s}{4m_e^2}- \frac{11}{6}}.
\label{eq:AngularDistribution}
\ea
The result for the total cross section, integrated in the
allowed kinematical limits is:
\ba
\sigma_{total}=
\frac{\sigma_0}{4}
\br{1-\frac{\beta^2}{3}}
\br{\ln \frac{s}{4m_e^2} - \frac{11}{6}}.
\label{eq:TotalCS}
\ea

The $\rho$-meson contribution can be included through the following
replacement:
\ba
    \sigma_0
    \to
    \tilde\sigma_0
    =
       \frac { \alpha s^2
    }
    {
     2^3\pi^3 3 \beta~}
    \brm{
        \frac{
            g_{\omega \pi\gamma^*}
            \brm{F_1^\omega\br{s}}
        }{s-M^2_\omega+iM_\omega\Gamma_\omega}
        +
        \frac{
            g_{\rho \pi\gamma^*}
            \brm{F_1^\rho\br{s}+F_2^\rho\br{s}}
        }{s-M^2_\rho+iM_\rho\Gamma_\rho}
	e^{i \phi_\rho}
    }^2,
\ea
where $\phi_\rho \approx 110^{\circ}$ \cite{Am08} is the relative phase of
$\rho$-meson contribution.

The main ingredients of the model are contained in the vertices: the strong form factors, in the vertex $VNN$ and the coupling constants $G_{V\pi\gamma^*}$. Following \cite{Ma00} (assuming its validity in the present range) $F_{1,2}^V\br{s}$ are parametrized as:
\ba
    F_1^V(s) = g_{Vpp}\frac{\Lambda_V^2-M_V^2}{\Lambda_V^2+s},
    \qquad
    F_2^V(s) = \kappa_V F_1^V\br{s},
 \label{eq:FFVpp}
\ea
where the constant $g_{Vpp}$ corresponds to the coupling
of the vector meson $V$ with protons. $\kappa_V$ plays the role of 'anomalous magnetic moment' of proton with respect to the coupling with the vector
meson $V$. $\Lambda_V$ is an empirical cut-off.

The quantity $G_{V\pi\gamma^*}$  is related to the
vertex $V(q)\to\pi(q_{\pi})\gamma^*(k)$ and it is in principle momentum dependent. It can be calculated in the framework of Nambu-Jona-Lasinio model, as illustrated in Fig. \ref{Fig:coupling} (black, solid line). A phenomenological parametrization, based on a monopole (dashed line) or dipole (dotted line) dependence on $q^2$ is also illustrated. The normalization $g_{V\pi\gamma}\br{0}$ can be derived from the radiative decay $V\to\pi\gamma$. One finds $g_{\omega\pi\gamma}=1.8$ ($g_{\rho\pi\gamma}=0.56$) for $\omega$($\rho$) meson respectively.

\begin{figure}
\begin{center}
\includegraphics[width=11cm]{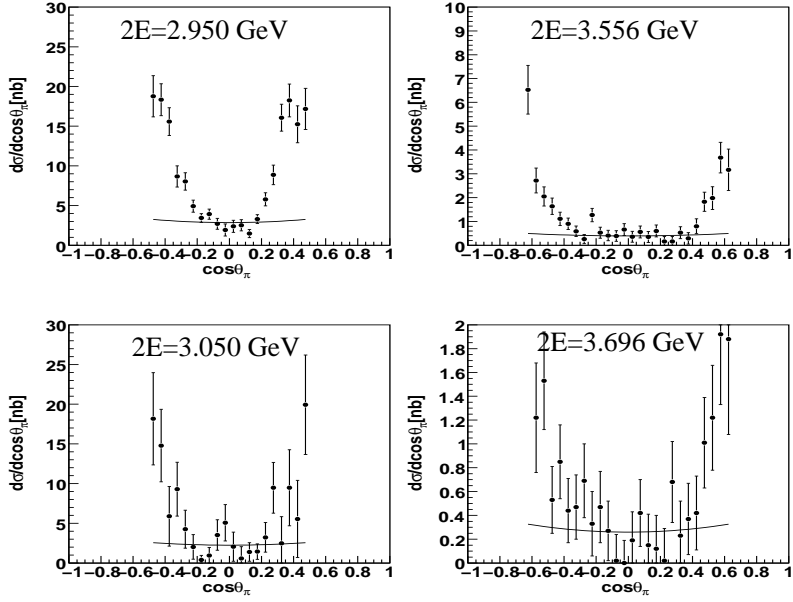}
\caption{\label{Fig:ang} Angular distributions for different values of the cms energy. The data are from \protect\cite{Arm97}, the line is a comparison with the model which applies at large angles.}
\end{center}
\end{figure}

The bidimensional plot of the cross section as a function of the CMS angle and energy of the pion is shown in Fig. \ref{Fig:bidim}. The angular and energy dependences are fixed by the model. The cross section is ploted in arbitrary units, due to the different possible choices for the vertex description. No data exist on reaction (\ref{eq:eqPPM}), but the reaction $\bar p +p\to\gamma+\pi^0$ was measured in the
region $2.911\le\sqrt{s}\le 3.686$ by the Fermilab E760 Collaboration \cite{Arm97}. For the latter reaction, few data exist on the cross section and the angular distributions. It is in principle possible to modify the present model for $\gamma$ production, and compare the calculation to  large angle data, where the present mechanism is expected to be dominant (Fig \ref{Fig:ang}). 

\section{The ratio of kaon to pion pairs in  $\bar p+p$ annihilation} 

Let us discuss the interest of measuring the ratio of $p\bar p$ annihilation into kaon and pion pairs near the threshold region, where the $S$ state dominates. The particular interest of these processes is related to the availability of high energy antiproton beams at PANDA (FAIR), in next future. The interesting kinematical region, the threshold region, can be reached in the collision of a $\bar p$ beam using the known mechanism of 'return to resonances', where the excess energy is carried by photon or jet emission.

The ratios:
$$R_s=\displaystyle\frac {(p\bar p)_{{\cal J}=0}\to K\bar K}{(p\bar p)_{{\cal J}=0}\to \pi^+\pi^-},~
R_p=\displaystyle\frac {(p\bar p)_{{\cal J}=1}\to K\bar K}{(p\bar p)_{{{\cal J}=1}}\to \pi^+\pi^-},$$
contain useful information on the reaction mechanisms, and may give evidence for the properties of the vacuum excitations. 

\begin{figure}
\begin{center}
\includegraphics[width=8cm]{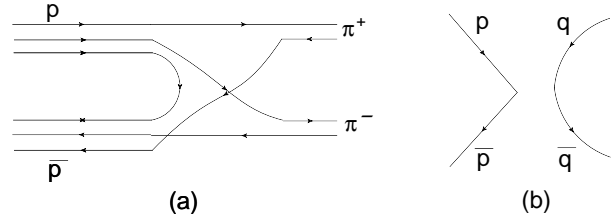}
\caption{\label{Fig:figev} (a) Feynman diagram for the reaction $ \bar{p} +p \to \pi^++\pi^-$ through quark rearrangement; (b) scheme for production of a pair of quarks through vacuum excitation.}
\end{center}
\end{figure}

A charged pion pair is expected to be produced at threshold from a $p\bar p$ triplet state, through rearrangement of the constituent quarks (see Fig. \ref{Fig:figev}a),  whereas kaon production is forbidden due to the absence of constituent strange quarks in the proton ground state.
However a kaon pair can be produced through a disconnected diagram (Fig. \ref{Fig:figev}b) where any pair of particles can be created from the excited vacuum. In this case, the singlet state, $^1S_0$ can produce a pair of current quarks, which, after interacting, become observable as mesons. The probability to produce a $u\bar u$ pair of current quarks is the same as for a $s\bar s$ pair of current quarks due to the structure of excited vacuum (EV).
Therefore the expectations are that $R_p\ll R_s\simeq 1$, that the angular distribution of kaon production is isotropic, whereas pions would follow an angular dependence driven by the dominant contribution of triplet state. 

The matrix element of the process $\bar p p \to \bar q q$ in singlet state, is proportional to $\bar u(p_-) v(p_+)$ (where $u,v$ are the spinors of the quark and antiquark, with four momentum $p_+$, $p_-$respectively):
\be
\left |M(\bar p p \to EV\to \bar q q)\right |^2\sim  Tr(\hat p_+-m_q)(\hat p_-+m_q)=8\beta_q^2 M^2,~q=u,d,s,
\ee
where $\beta_q=\sqrt{1-m^2_q/E_q^2}$,  $m_u=m_d=280$ MeV, $m_s=400$ MeV, and  $E_u=E_s=M$.

The yield of pion and kaon pair production should be corrected by the phase volume ratio $\phi_{\pi}/\phi_K=\beta_{\pi}/\beta_K$. Finally
\be 
R=\frac{f(K^+K^-)}{f(\pi^+\pi^-)}=\frac{1/2}{3+1/2}\frac {\beta_K}{\beta_{\pi}}
\left ( \frac {\beta_s}{\beta_{u}}\right )^2=0.108
\label{eq:eqYY}
\ee
to which we attribute an error of $5\%$ related to the constituent quark masses.
This quantity may be compared to the experimental value $R=0.108\pm 0.007$ \cite{Ab94}. The present estimate is based on the statement that pion and kaon pairs are produced 'democratically' from the vacuum excitation, whereas kaon production is forbidden in singlet $S$ state.
In the kinematical region accessible to PANDA (FAIR), in principle, the threshold region can be reached even if the $\bar p$ beam momentum is as high as 15 GeV,  using the 'return to resonances' mechanism, where the excess energy is carried, for example, by a hard real or a virtual photon. Using conservation laws and on-mass condition for a protonium ($p\bar p$ bound state, $B$ of momentum $P$):
\ba
&\bar p(p_1)+p(p_2)&\to\gamma(k)+B(P),~k^2=0,~P^2=4M^2,~p_1^2=p_2^2=M^2;
\nonumber\\
&\bar p(p_1)+p(p_2)&\to\gamma^*(k)+B(P),~k^2=M_X^2\gg 0
\label{eq:eqreac}
\ea
one obtains, for the case of real photon emission:
\be
(p_1 + p_2-k)^2=4M^2,~\omega=\displaystyle\frac{E-M}
{ 1+({E}/{M})(1-\beta\cos\theta)},
~\beta=\sqrt{1-\displaystyle\frac{M^2}{E^2}}
\label{eq:eqg}
\ee
where $\theta=\widehat{\vec k \vec {p_+}} $ and $\beta$  are the photon emission angle and the antiproton velocity, and  $E$ is the energy  of the beam (the direction of the beam is taken as the $z-$ axis). 
\section{Conclusions}
We have given two examples of exclusive processes induced by antiproton beams in the GeV range. The annihilation of $\bar p +p\to e^+ +e^-+\pi^0$ contains information on nucleon electromagnetic form factors in the unphysical region. The mechanism of production through s-channel exchange of a vector meson is especially important at large angles and should be studied as a physical background, which contains information on  meson nucleon interaction and meson decays.

The ratio of kaon pair to pion pair production at threshold contains information on the reaction mechanism, and could give evidence for excitation of the physical vacuum.
It is shown that a good agreement with existing data on the ratio of the yields of charged kaon to pion pairs is obtained taking into account not only quark rearrangement in the incident hadrons, but also vacuum excitations.

Acknowledgments are due to G.I Gakh, V. Bytev and to the IPNO-PANDA group for a careful reading of the manuscript and for useful discussions. This work was partially supported by GDR n.3034 'Physique du Nucl\'eon' (France).

\end{document}